\documentclass[a4paper,fleqn,usenatbib,useAMS]{mnras}
\usepackage{epsfig,amssymb,amsmath,rotating,lineno,graphicx,hyperref}
\usepackage[T1]{fontenc}
\usepackage{ae,aecompl}
\usepackage{xcolor}
\newcommand\asat{{\it AstroSat}}
\newcommand\xte{{\it RXTE}}

\newcommand\nicer{{\it NICER}}

\newcommand\xmm{{\it XMM-Newton}}
\newcommand\swift{{\it SWIFT}}
\title[Timing study of GX~340+0]{\asat{} and \nicer{} timing view of the Z$-$type Neutron Star X-ray binary GX~340+0}

\author[Pahari et al.]{Mayukh Pahari$^{1}$, Shree Suman$^{1}$, Yash Bhargava$^{2}$, Alexander Weston$^{3}$, Liang Zhang$^{4}$, 
\newauthor Sudip Bhattacharyya$^{2}$, Ranjeev Misra$^{5}$, Ian McHardy$^{3}$ \\
$^{1}$ Department of Physics, Indian Institute of Technology, Hyderabad, Kandi, Sangareddy 502285, India \\
$^{2}$ Department of Astronomy and Astrophysics, Tata Institute of Fundamental Research, 1 Homi Bhabha Road, Colaba,
Mumbai 400005, India \\
$^{3}$ School of Physics \& Astronomy, University of Southampton, Highfield, Southampton SO17 1BJ, UK \\
$^{4}$ Key Laboratory of Particle Astrophysics, Institute of High Energy Physics, Chinese Academy of Sciences, Beijing 100049, China \\
$^{5}$ Inter-University Centre for Astronomy and Astrophysics, Pune, 411007, India}

\begin{document}

\pagerange{\pageref{firstpage}--\pageref{lastpage}} \pubyear{2024}

\maketitle

\label{firstpage}

\begin{abstract}
The timing properties of the Z-type low-mass X-ray binaries provide insights into the emission components involved in producing the unique Z-shaped track in the hardness-intensity diagrams of these sources. In this work, we investigate the \asat{} and \nicer{} observations of the GX~340+0 covering the complete `Z'-track from the horizontal branch (HB) to the extended flaring branch (EFB). For the first time, we present the Z-track as seen in soft X-rays using the \asat{}/SXT and \nicer{} (the soft colour is defined as a ratio of 3--6~keV to 0.5--3~keV). The shape of the track is distinctly different in soft X-rays, strongly suggesting the presence of additional components active in soft X-rays. The detailed timing analysis revealed significant quasi-periodic oscillation throughout the HB and the normal branch (NB) using LAXPC and the first \nicer{} detection of 33.1 $\pm$ 1.1~Hz horizontal branch oscillation (HBO) in 3--6~keV. The oscillations at the HB/NB vertex are observed to have higher frequencies (41--52~Hz) than the HB oscillations (16--31~Hz) and NB oscillations (6.2--8~Hz) but significantly lower rms ($\sim$1.6\%). The HB oscillation is also limited to the energy range 3--20~keV, indicating an association of HBO origin with the non-thermal component. It is also supported by earlier studies that found the strongest X-ray polarisation during HB.

\end{abstract}


\section{Introduction}
Neutron star low mass X-ray binaries (NSXBs) can be divided into two main classes, atoll and Z, depending on the shape of a Hardness-Intensity Diagram (HID) or a Colour-Colour Diagram (CCD) \citep{1989A&A...225...79H}.  
`Z'-type sources are divided into two groups based upon the length of branches: `Cyg-like' sources, which include Cyg~X-2, GX~340+0, GX~5-1 and `Sco-like' sources, which include Sco~X-1, GX~17+2, and GX~349+2 \citep{ku94,ho10}.
`Z' sources draw out three main sections on the CCD/HID as they evolve with time: the horizontal branch (HB), the normal branch (NB), and the flaring branch (FB). Sometimes, a fourth branch, which we shall hence refer to as the extended flaring branch (EFB), is also observed. The HB and NB are joined by an upper vertex (also called a hard apex), whilst the NB and FB are joined by a lower vertex (also called a soft apex).
The sources are observed to spend an arbitrary duration of time at a particular HID position but have always shown a motion along the Z-track (without jumping across different branches).  However, the time spent by the source in EFB is substantially shorter than the HB, NB, or FB. The HB is the hardest spectral state \citep{se13,bh23} whilst the FB is the softest one among the three branches. 
The simple hypothesis of mass accretion rate $\dot{M}$ increasing monotonically with the source moving from the HB $\rightarrow$ NB $\rightarrow$ FB \citep{1986ApJ...306L..91P} is now widely believed not to be the explanation for the Z-track movement. This is because the X-ray intensity decreases as the source moves along the NB, opposite to what one would expect from an increasing $\dot{M}$. More complex versions of this theory exist \citep{2010A&A...512A...9B, 2012A&A...546A..35C}. The transition between different branches, particularly HB to NB, was found to have a connection to the launch of radio jets and outflows \citep{mi06}; in particular, Sco~X-1 showed an ultra-relativistic flow associated with the state transition \citep{mo19}.
Multi-wavelength variability studies of Z sources tentatively indicate that they share the same physical processes of highly accreting BHs \citep{mu14,vi23}.
Although no clear consensus exists on what physical models best describe the spectral evolution along different branches, spectra are usually described as a combination of soft thermal emission from the accretion disc, a black-body emitting component, known as the boundary layer, and a hard, non-thermal comptonised emission-dominated component \citep{li07,ho07,li09,ho10,li12,bh23}. 

With the launch of Imaging X-ray Polarimetry Explorer (IXPE), more robust details of physical processes in the vicinity of NS surface in `Z' type NSXBs have been studied, e.g., during the 2022 outburst of XTE J1701-462, \citet{co23} measured an average 2-8~keV polarisation degree of 4.6 $\pm$ 0.4\% which is primarily due to the very hot black-body emission from the boundary layer. Polarisation is found to be strongest in the HB and weakest in the FB in XTE J1701-462. GX 5-1 also show stronger X-ray polarisation in HB ($\sim$3.7\%) compared to the NB and FB ($\sim$1.8\%) \citep{fa23}. Using \textit{IXPE}, \textit{NICER} and \textit{INTEGRAL}, \citet{fe23} studied the spectro-polarimetric properties of Cyg~X-2 and observed that the comptonised emission has a polarisation degree of 4.0 $\pm$ 0.7\% while the polarisation angle is aligned with the direction of the radio jet. 

In addition to spectroscopic variability, NSXBs often display distinct quasi-periodic oscillations (QPOs) and broad Lorentzian components in power density spectra (PDS). They come in many flavours with a very broad range of frequencies see \citep[see][and references therein]{jo98,jo00,2004astro.ph.10551V}. QPOs are often present on the HB and upper NB regions \citep{va88,pe91,jo98,jo00,sr11,bh23} and in a subset of Z-sources, also in the FB (e.g., $\sim$26~Hz in Cyg~X-2,  \citep{wi98}, $\sim$15 Hz in Sco~X-1 \citep{ca06} and $\sim$7-20 Hz in GX~17+2 \citep{pe91}). Using \xte{} data of GX~340+0, \citet{jo00} showed that HBO increases monotonously from 20 to 50~Hz and saturates afterwards, while their rms amplitude continuously decreases from 10\% to 2\%. 
Although a single model cannot explain the origin of HBOs, the magnetospheric beat frequency (MBF) model \citep{al85,la85} is preferred over the Lens-Thirring precession (LT) model \citep{st98} since it can predict the increase in HBO frequencies with the increasing accretion rate. In the case of GX~340+0, \citet{jo98}, showed that the LT predicted HBO is smaller by a factor of $\sim$3 than that observed during the \xte{} campaign. 

GX~340+0 is identified as a luminous, low-mass neutron star X-ray binary, showing a `Z'-shaped track in the HID \citet{1989A&A...225...79H}. 
The X-ray timing characteristics and power spectral attributes of GX~340+0 were previously elucidated by various studies. \citet{va88,ku96} examined the data acquired from the \textit{EXOSAT} satellite. \citet{pe91} analysed data from the \textit{Ginga} satellite, while \citet{jo98} used data from the \xte{} satellite. Notably, \citet{pe91} and \citet{jo98} both noted the presence of EFB in GX~340+0 following the FB in the HID.

When GX~340+0 resides on the HB or the upper portion of the NB in the HID, it exhibits QPOs with frequencies ranging from 20 to 50~Hz. These oscillations are called horizontal-branch oscillations or HBOs, as detailed by \citet{pe91,ku96,jo98}. Additionally, \citet{ku96} and \citet{jo98} reported the detection of second harmonics of these HBOs, with frequencies in the ranges of 73$-$76~Hz and 38$-$69~Hz, respectively. During HB \citet{jo00} found strong paired kHz QPOs whose frequencies evolved by $>$300~Hz within a day of observation.
In the middle of the NB, \citet{va88} identified normal-branch oscillations (NBOs) with a frequency of 5.6~Hz.  

 The unique shape of the Z-track, coupled with the various temporal features associated with different branches, proves to be an enigma. One of the key probes to understand HBOs and NBOs is investigating the source evolution in soft X-rays, which have been sparse due to the intrinsic brightness of the source leading to pile-up effects in soft X-ray instruments like \xmm{} and \swift{}/XRT. Only recently, \nicer{} has been used to investigate the timing properties of Cyg~X-2 and Sco~X-1 \citep{ji23}, and NBOs down to 0.4~keV are detected. With \asat{}/SXT coupled with \asat{}/LAXPC uniquely provides broadband information of `Z' sources both in soft and hard X-ray bands \citep{bh23}. 

In this work, we investigate the timing properties of the GX~340+0 using \asat{} and \nicer{} observations of the source during different epochs. Combining 2017 and 2018 epochs of \asat{} observations consisting of $\sim$202~ks of data, we have obtained the complete `Z' track of the source using the hardness definition of 10--20~keV to 6--10~keV. Using simultaneous \asat{}/SXT data, we depict the soft band HID of any `Z' source for the first time. The behaviour is qualitatively similar to that of the hard band `Z' track observed using LAXPC data. However, the slope of the SXT HB track is opposite to that observed from LAXPC. Using \nicer{} data, we have detected HBO at $\sim$33~Hz down to 0.5~keV, the first HBO detection using \nicer{} observation. Combining PDS properties of \nicer{} and \asat{}, we found that energy-dependent HBO rms amplitude increases with the frequency. However, at the HB/NB apex, the rms fall significantly at the QPO frequency of 41--52~Hz. In addition, the 41--52~Hz QPO frequencies increase with the photon energies, while they remain constant at 16--33~Hz. Therefore, 41-52 Hz HB/NB vertex  QPOs may have different origins than 16-33 Hz QPOs which are mostly confined to the horizontal part of the `Z' track. 

\begin{figure*}
\centering
\includegraphics[scale=0.69]{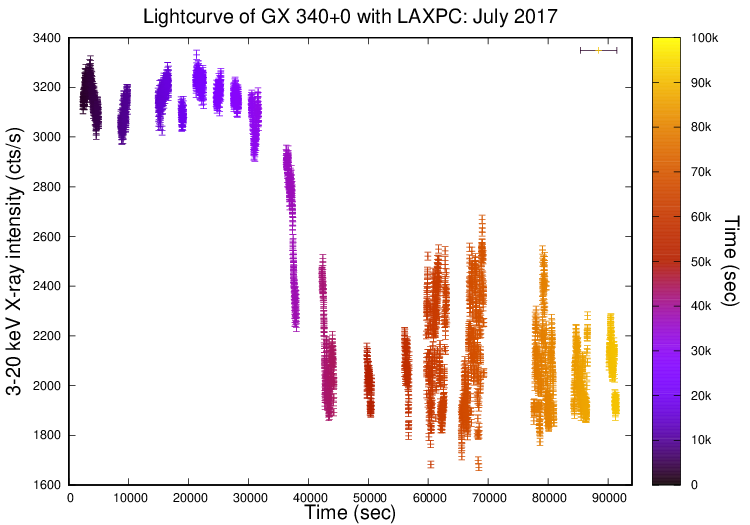}
\includegraphics[scale=0.69]{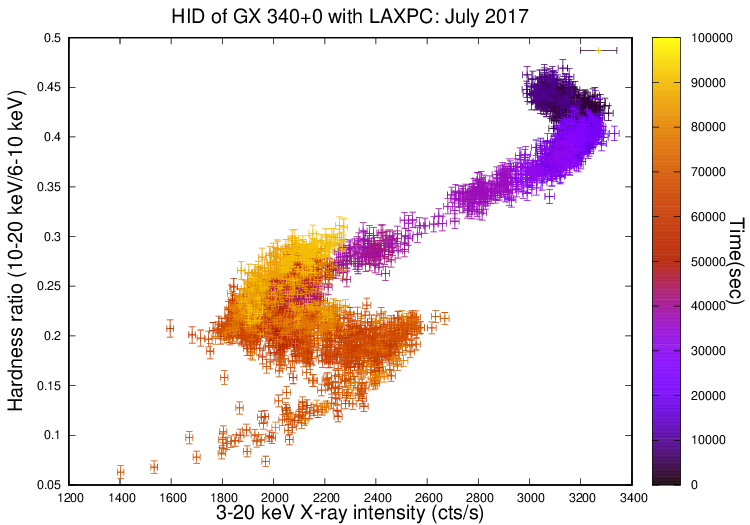}
\caption{3--20~keV background-subtracted lightcurve from GX~340+0 combining LAXPC10 and 20 are shown for the observation made in the July 2017 epoch (left panel). The hardness intensity diagram (HID) of the same epoch is shown in the right panel, where the hardness is defined as the ratio of count rate in 10--20~keV and 6--10~keV, and the X-ray intensity is defined as the count rate in 3--20~keV. The same colour is used in both panels to represent the time evolution of the sources in the HID and the light curve. }
\label{light-hid-2017}
\end{figure*}

\section{Observations and Data Reduction}
\subsection{AstroSat}
GX~340+0 data were obtained via India's multi-wavelength observatory, \asat{} \citep{si14}, using simultaneous observations from both large area X-ray proportional counter (LAXPC) and soft X-ray telescope (SXT) instruments. We have used two epochs of observations provided in Table~\ref{obs}. The first epoch started on 30th July 2017 at 16:39:55~UT and lasted for $\sim$93~ks covering nearly 17 \asat{} orbits (9939-9957; except 9940, 9950 and 9954). The second epoch started on 06 May 2018 at 12:52:35 UT and lasted for $\sim$112~ks covering 20 \asat{} orbits (14083-14102). Details of orbit-wise start time and exposures are provided in Table~\ref{orbit}.

\begin{table*}
\centering
\caption{Details of \asat{} and \nicer{} observations used in the present work }
\label{obs}
\begin{tabular}{ccccc}
\hline
Mission & Start time & Instrument  & Observation & Exposure \\
  & (YY-MM-DD) & & ID & (ks) \\
\hline
AstroSat & 17-07-30   & LAXPC & $G07\_016T01\_9000001420$ & 92.4 \\
AstroSat &  & SXT & $G07\_016T01\_9000001420$  &  48.8 \\
AstroSat & 18-05-06  & LAXPC & $G08\_022T01\_9000002078$ & 111.8 \\
AstroSat & & SXT & $G08\_022T01\_9000002078$ & 77.2 \\
NICER & 18-03-24 & XTI & 1108010105 & 13.3 \\
NICER & 18-08-13 & XTI & 1108010108 & 29.8 \\
NICER & 18-08-17 & XTI & 1108010111 & 24.8 \\
\hline
\end{tabular}
\end{table*}

\begin{table}
\centering
\caption{Orbit-wise details Details of \asat{} observations }
\label{orbit}
\begin{tabular}{cccc}
\hline
& 30-31st July 2017 & & \\
\hline
Orbit & Date & Start time (UT) & Exposure  \\
Number  & & (hh:mm:ss) & (sec)  \\
\hline
09939 & 30-07-17 & 15:47:14 & 4369.0 \\
09941 & 30-07-17 & 16:42:23 & 7247.0 \\
09942 & 30-07-17 & 18:11:29 & 8017.6 \\
09943 & 30-07-17 & 19:58:31 & 7807.4 \\
09944 & 30-07-17 & 21:49:14 & 7444.6 \\
09945 & 30-07-17 & 23:34:33 & 7440.3 \\
09946 & 31-07-17 & 00:52:43 & 9172.0 \\
09947 & 31-07-17 & 02:53:29 & 8074.3 \\
09948 & 31-07-17 & 04:24:01 & 8971.1 \\
09949 & 31-07-17 & 06:17:42 & 8477.5 \\
09951 & 31-07-17 & 08:01:39 & 14914.3 \\
09952 & 31-07-17 & 11:56:40 & 6947.1 \\
09953 & 31-07-17 & 13:40:38 & 7083.3 \\
09955 & 31-07-17 & 15:16:28 & 7549.9 \\
09957 & 31-07-17 & 17:04:12 & 1431.6 \\
\hline
& 06-07 May 2018 & & \\
\hline 
14083 & 06-05-18 & 13:00:45 & 8136.0 \\
14084 & 06-05-18 & 14:47:56 & 7871.4 \\
14085 & 06-05-18 & 16:34:28 & 7712.5 \\
14086 & 06-05-18 & 18:06:00 & 8481.9 \\
14087 & 06-05-18 & 19:49:41 & 8643.1 \\
14091 & 06-05-18 & 21:33:22 & 27664.3 \\
14092 & 07-05-18 & 04:59:45 & 6926.9 \\
14093 & 07-05-18 & 06:40:01 & 7225.7 \\
14095 & 07-05-18 & 08:12:16 & 8212.3 \\
14096 & 07-05-18 & 09:51:26 & 8585.9 \\
14097 & 07-05-18 & 11:31:32 & 8742.7 \\
14098 & 07-05-18 & 13:36:49 & 7365.6 \\
14099 & 07-05-18 & 15:21:06 & 7211.8 \\
14100 & 07-05-18 & 16:55:52 & 7839.2 \\
14101 & 07-05-18 & 18:22:53 & 8803.8 \\
14102 & 07-05-18 & 20:24:26 & 7987.7 \\
\hline
\end{tabular}
\end{table}

\subsubsection{LAXPC}
LAXPC \citep{2016SPIE.9905E..1DY,2017ApJS..231...10A} consists of 3 independent but identical detectors giving a collecting area of $\sim$6000~cm$^{2}$ at 15~keV with an operational energy range of 3-80~keV. The detectors have a dead time of about 42~$\rm{\mu}$s. 

The LAXPC data were reduced using \textsc{LaxpcSoft v21June2023}\footnote{\url{http://astrosat-ssc.iucaa.in/uploads/laxpc/}} with suitable response files for corresponding LAXPC units. Cleaned event files were created along with good time intervals (GTIs) for each satellite orbit using LAXPC10 and LAXPC20 units. Due to poor data quality and calibration issues, data from LAXPC30 is excluded from all further analyses. GTIs were cleaned further to remove segments with poor data quality caused by telemetry losses or other factors like the initial and the final 100~s of transition in and out to the South Atlantic Anomaly (SAA) region, respectively. Such a screening accounts for nearly 5\% of the total effective exposure. Lightcurves in 6-10~keV, 10-20~keV, and 3--20~keV energy bands are extracted using the cleaned GTIs, and hardness intensity diagrams are computed. 

\subsubsection{SXT}

 \asat{}/SXT is a focusing telescope utilizing a charge-coupled device capable of X-ray imaging in the energy range of 0.3-7.0~keV with medium resolution \citep{2016SPIE.9905E..1ES,2017JApA...38...29S}. We have used data from orbits the same as LAXPC. SXT data were reduced using \textsc{SXTPIPELINE v1.4b}\footnote{\url{http://www.tifr.res.in/~astrosat_sxt/sxtpipeline.html}}, and \textsc{xselect v2.4g}. All orbits were concatenated together and cleaned before a 12-arcmin radius region was selected for use as the source region. Light curves were extracted in the energy range of 0.5--3~keV and 3--6~keV, respectively. The total good time exposure for the 2017 and 2018 epochs are 78~ks and 119~ks, respectively.

\subsection{NICER}
The \nicer{} X-ray Timing Instrument \citep[XTI,][]{ge16} consists of 56 co-aligned X-ray concentrator optics working in the 0.2-12~keV band \citep{pr12}. The 52 operating detectors collectively provide an effective area of 1900 cm$^2$ at 1.5~keV, with an energy resolution of $\sim$100 eV. As part of its main science program, \nicer{} has extensively monitored the `Z' source GX~340+0 during 2018 and observed the source at 11 different epochs. In this work, we have used the three longest observations for which data are useful to study different parts of the `Z' track. Observation details are provided in Table~\ref{obs}. \nicer{} data are processed using {\tt nicerl2} pipeline in {\tt Heasoft 6.31.1} using the \nicer{} version of {\tt 2022-12-16\_V010a}, and CALDB version of {\tt xti20221001}. Standard filtering criteria are adopted: we selected data collected with a pointing offset of less than 54 arcsec, more than 40$^\circ$ away from the bright Earth limb and 30$^\circ$ away from the dark Earth limb, and outside the South Atlantic Anomaly. We then constructed space weather model-predicted background lightcurves and subtracted from source+background lightcurves with 8~s time resolution in the 12--15~keV and 6--12~keV energy bands. Whenever the source light curve had a rate greater than 1 cts/s in 12-15 keV,  we observed a correlated increase in the 0.4- 12 kV rate. We, therefore, attributed those epochs to high-background intervals and removed them from our analysis.  About 2.5~ks of high-background exposure was removed in this way. After filtering, we were left with approximately 26~ks worth of good time exposure. With customized GTI files, lightcurves are extracted in the energy ranges of 0.5--3~keV, 3--6~keV, 6--10~keV and 0.5--10~keV respectively using {\tt xselect} in {\tt HEASOFT 6.31.1}. 

\section{LAXPC analysis and results}

A 3--80~keV 10~s binned light curve was extracted by combining {\tt LAXPC10} and {\tt LAXPC20} observations and using \textsc{LaxpcSoft}. Light curves in different energies were obtained along with the PDS. The PDS was fitted with \textsc{XSpec v12.13.0c} included in \textsc{HEASOFT v6.31.1}. Unless otherwise stated, analysis of GX~340+0 data involved excluding photons of energy greater than 30~keV due to counts being similar to the background.
 
\subsection{Timing Analysis and Results}

\begin{figure*}
\centering
\includegraphics[scale=0.69]{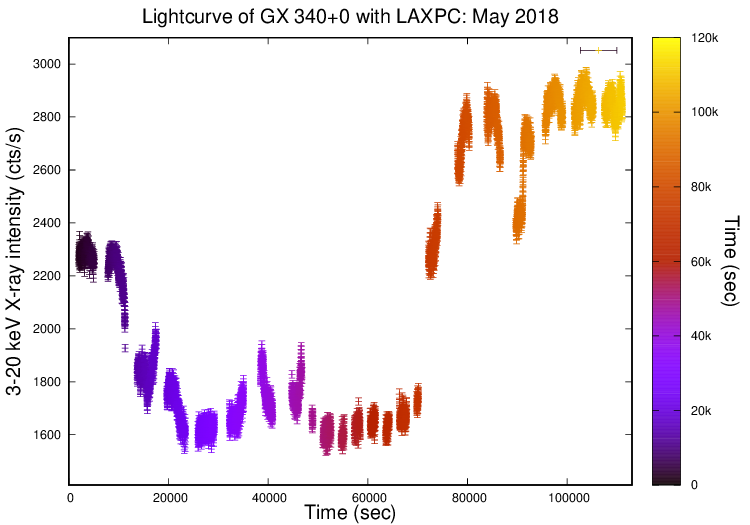}
\includegraphics[scale=0.69]{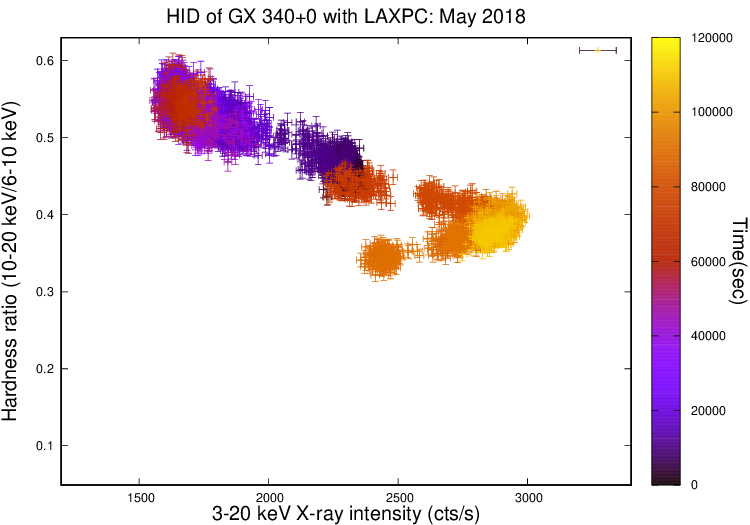}
\caption{3--20~keV background-subtracted lightcurve from GX~340+0 combining LAXPC10 and 20 are shown for the observation made in the May 2018 epoch (left panel). The hardness intensity diagram (HID) of the same epoch is shown in the right panel, where the hardness is defined as the ratio of count rate in 10--20~keV and 6--10~keV, and the X-ray intensity is defined as the count rate in 3--20~keV. The same colour bar is used in both panels to represent the time evolution of the sources in the HID and the light curve.}
\label{light-hid-2018}
\end{figure*}

3--20~keV background-subtracted lightcurve combining LAXPC10 and LAXPC20 is shown in the left panel of Figure~\ref{light-hid-2017} during 2017 epoch combining data from 17 orbits. A large change in X-ray intensity by a factor of $\sim$1.5-2 starting from $\sim$30~ks was observed, along with a strong variability in the count rate during 60ks-80ks of the observation.
Hardness intensity analysis was performed to understand the nature of the X-ray intensity variation. Because GX~340+0 is a known Z-source, it was believed that there would be a statistically significant change in hardness along with the drop in the count rate. The hardness ratio is defined as the ratio of background-subtracted count rate in 10--20~keV and 6--10~keV, respectively, while the X-ray intensity was defined as the X-ray count rate in the 3--20~keV energy range. This definition is chosen to be consistent with \asat{}/LAXPC hardness analysis of GX~340+0 presented in \citet{bh23}. The resulting HID is shown in the right panel of Figure~\ref{light-hid-2017}. The HID shows a part of HB, HB/NB vertex, NB, FB, and extended flaring branch (EFB). 
The colours in both panels indicate the observation time, therefore indicating the time evolution of the HID during the 2017 epoch. 

To obtain the complete `Z' track of GX~340+0, a similar exercise has been performed for the LAXPC observation of GX~340+0 during the second epoch in 2018, covering 20 \asat{} orbits.
3--20~keV background-subtracted lightcurve combining LAXPC10 and LAXPC20 is shown in the left panel of Figure~\ref{light-hid-2018} during 2018 epoch. A sudden rise in X-ray flux by a factor of $1.5-2$ is observed around $\sim$72~ks. 
The right panel of Figure~\ref{light-hid-2018} shows complete HB along with the transition to NB via HB/NB vertex. The colour bars used in Figure~\ref{light-hid-2018} are kept similar to those used in Figure~\ref{light-hid-2017}.

Therefore, both epochs together show the entire `Z' track of GX~340+0 with \asat{}/LAXPC. The colours indicate that the `Z' track is continuously tracked by the source from HB $\rightarrow$ NB $\rightarrow$ FB $\rightarrow$ EFB. After that, the source returns to the FB again. During the 2018 epoch, the source evolution is non-monotonous and makes a back-and-forth flux transition within HB. Then, the sudden rise in the X-ray intensity corresponds to the transition from HB $\rightarrow$ NB via the HB/NB vertex.

To conduct further timing analysis, we have extracted PDS using lightcurves of each orbit from both epochs and fit with power-law and Lorentzians to find out and quantify noise and QPO components from each PDS fitting.  

\subsection{Power spectral analysis and results}

\begin{figure*}
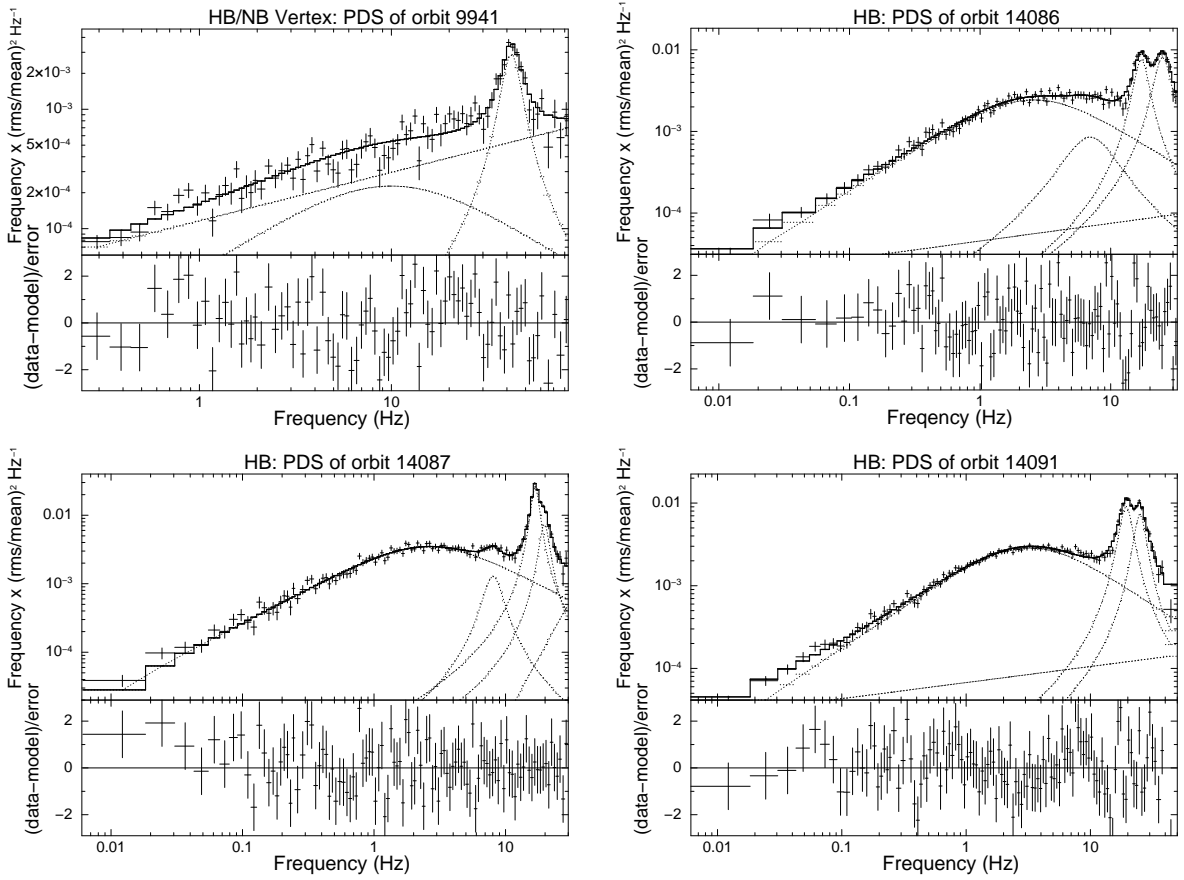

\includegraphics[scale=0.3,angle=-90]{fig3a.ps}
\includegraphics[scale=0.3,angle=-90]{fig3b.ps}
\includegraphics[scale=0.3,angle=-90]{fig3c.ps}
\includegraphics[scale=0.3,angle=-90]{fig3d.ps}
\caption{rms-normalised and Poisson-noise subtracted best-fit power density spectra (fitted with combinations of the power-law and Lorentzians) from HB and HB/NB vertex are shown in 3-30~keV along with their residuals for the orbit number 9941 (top left), 14086 (top right), 14087 (bottom left) and 14091 (bottom right) respectively. QPOs above 10~Hz are observed in all cases, and double QPOs are observed in a couple of cases. }
\label{hfpds}
\end{figure*}

For each orbital data from \asat{}, we have extracted rms-normalized and Poisson-noise subtracted power PDS in the 3--30~keV energy range using good time intervals after removing data gaps and count drops due to telemetry losses. Each PDS is fitted separately using a combination of power-law and multiple Lorentzians in the frequency range of 0.05--70~Hz. From the best-fit model components, total fractional rms, QPO frequencies, and their fractional rms are calculated. Fitted parameter values for different orbits during the 2017 and 2018 epochs are provided in Table~\ref{2017-pds-table} and Table~\ref{2018-pds-table}, respectively. Bestfit PDS from four different orbits, 9941, 14086, 14087, and 14091, are shown in the top left, top right, bottom left, and bottom right panels of Figure~\ref{hfpds} respectively, along with their residuals.

We found a strong QPO in the GX~340+0 in the frequency range of $\sim$41--52~Hz during both epochs. Comparing Figure~\ref{light-hid-2017} with Table~\ref{2017-pds-table} and Figure~\ref{light-hid-2018} with Table~\ref{2018-pds-table}, we noted that the occurrence of high-frequency QPOs around $\sim$41--52~Hz is always associated with the HB/NB vertex during both epochs. When the source made a transition to either HB or NB from the vertex, $\sim$ 50~Hz QPOs were not observed. One such QPO is shown in the top left panel of Figure~\ref{hfpds}. Using the data from the 2020 epoch, \citet{bh23} found QPOs at $\sim$43.5~Hz and $\sim$48.5~Hz precisely during the HB/NB vertex transition in GX~340+0. Detection of $\sim$42--52~Hz QPOs exclusively during HB/NB vertex during three different epochs of observations in 2017, 2018, and 2020 implies that such a feature could be a landmark for the transition from the horizontal branch to the normal branch in GX~340+0. 

HBOs are observed within the frequency range of 16--31~Hz during several orbits of the 2018 epoch where the HID is dominated by HB. However, no QPOs in a similar frequency range have been observed from PDS during the 2017 epochs of observations. Therefore, such a range of QPO frequencies may be exclusively associated with HBOs. HBOs in 16--31~Hz are also reported earlier \citep{jo00,bh23}. An example PDS with the HBO at $\sim$16.9~Hz is shown in the bottom left panel of Figure~\ref{hfpds}. During two orbits, we detect double-peaked QPOs: $\sim$18.5~Hz and $\sim$24.9~Hz during the orbit 14086 and $\sim$18.7~Hz and $\sim$24.7~Hz during the orbit 14091. They are shown in the top right and bottom right panels of Figure~\ref{hfpds}.

Our observations did not show any significant kHz QPOs. This is likely due to the fact that they are rare compared to HBOs, and our observation missed a large portion of the HB during the 2017 epoch, where all kHz QPOs are typically observed. 

\begin{table}
    \centering
    \caption{Power density spectral properties for the observation of July 2017 are provided orbit-wise. Observed QPOs and their fractional rms and the total PDS rms in the frequency range of 0.05--70~Hz are provided. 41--52~Hz QPOs have quality factors (Q, defined as $\nu$/$\delta \nu$) of 3.2--4.9, while 6.3--8.1~Hz NBOs have QFs in the range of 2.6--3.8.    }
    \label{2017-pds-table}
    \begin{tabular}{c|c|c|c}
    \hline
         Orbit & Total rms & Observed QPO & Observed QPO\\
         Number & (\%) & Frequency (Hz) &  rms (\%)\\
    \hline
         09939 & $2.88^{+1.16}_{-0.93}$ & $46.7^{+1.2}_{-1.5}$ & $1.79^{+0.43}_{-0.29}$\\ 
         09941 & $1.51^{+0.22}_{-0.18}$ & $45.2^{+1.1}_{-1.3}$ & $1.43^{+0.42}_{-0.16}$ \\
         09942 & $10.16^{+0.41}_{-0.28}$ & $7.82\pm0.07$ & $1.93^{+0.10}_{-0.08}$\\ 
         09943 & $1.88^{+1.16}_{-1.15}$ & $51.4^{+1.5}_{-1.4}$ & $1.09^{+0.19}_{-1.15}$ \\
         09944 & $6.54^{+0.16}_{-0.15}$ & $8.03^{+0.23}_{-0.16}$ & $1.02^{+0.17}_{-0.13}$\\ 
         09945 & $4.17^{+0.81}_{-0.27}$ & $7.48^{+0.72}_{-3.81}$ & $1.07^{+1.51}_{-0.29}$\\
         09946 & $2.04^{+0.19}_{-0.15}$ & --- & --- \\ 
         09947 & $1.03\pm0.17$ & --- & ---\\
         09948 & $1.66^{+0.28}_{0.15}$ & --- & ---\\ 
         09949 & $8.63^{+1.43}_{-0.33}$ & $6.25\pm0.12$ & $1.94^{+0.12}_{-0.29}$\\
         09951 & $8.91^{+0.22}_{-0.21}$ & $6.21^{+0.08}_{-0.07}$ & $1.50^{+0.10}_{-0.11}$\\ 
         09952 & $15.17^{+0.92}_{-1.04}$ & $6.29\pm0.05$ & $2.76\pm0.07$\\
         09953 & $1.58^{+0.49}_{-0.21}$ & --- & ---\\ 
         09955 & $2.00^{+0.26}_{-0.22}$ & --- & ---\\
         09957 & $1.31^{+1.01}_{-0.22}$ & --- & --- \\ 
    \hline
    \multicolumn{4}{l}{$^f$ parameter fixed at this value during fitting}
    \end{tabular}
     
\end{table}

\begin{table*}
\centering
\caption{Power density spectral properties for the observation of May 2018 are provided orbit-wise. Observed QPOs and their fractional rms and total rms in the frequency range of 0.05--70~Hz are provided. 41--43~Hz QPOs have quality factors (QFs) of 3.8--4.2, while 16.2--33.1~Hz HBOs have QFs in the range of 3.3--7.8.}
\label{2018-pds-table}
\begin{tabular}{cccc}
\hline
Orbit & Total rms  & QPO Frequency & QPO rms\\
 Number & (\%) & (Hz) & (\%) \\
\hline
14083&0.37$^{+0.09}_{-0.13}$ &--&-- \\
14084& 1.99$^{+0.26}_{-0.23}$ & 41.62$^{+0.47}_{-0.68}$ & 1.02$^{+0.27}_{-0.43}$\\
14085&19.34$^{+1.16}_{-0.91}$ & 28.13$^{+0.15}_{-0.14}$&7.45$^{+0.19}_{-0.18 }$\\
14086 & 19.32$^{+1.39}_{-2.31}$ & 18.52$^{+0.06}_{-0.09}$ & 5.95$^{+0.19}_{-0.21}$\\
&  & 24.93$^{+0.17}_{-0.19}$& 7.68$^{+0.74}_{-0.02}$ \\
14087& 17.22$^{+0.81}_{-0.70}$ & 16.92$^{+0.05}_{-0.04}$& 3.68$^{+0.22}_{-0.59}$\\
14091& 20.03$^{+0.16}_{-0.15}$ & 18.86 $^{+0.06}_{-0.06}$ & 5.32$^{+0.08}_{-0.08}$\\
&  & 24.66$^{+0.10}_{-0.10}$& 5.28$^{+0.10}_{-0.10}$\\
14092& 17.61$^{+0.43}_{-0.40}$ & 20.72$^{+0.07}_{-0.07}$ & 7.74$^{+0.17}_{-0.20}$\\
14093& 17.27$^{+0.44}_{-0.47}$ & 22.44$^{+0.10}_{-0.10}$ & 7.52$^{+0.15}_{-0.13}$\\
14095& 13.23$^{+0.30}_{-0.32}$ & 30.70$^{+0.11}_{-0.10}$& 5.19$^{+0.11}_{-0.15}$\\
14096& 6.07$^{+0.36}_{-0.42}$ & 42.35$^{+1.03}_{-0.65}$& 2.39$^{+0.19}_{-0.17}$\\
14097 & 0.55$^{+0.06}_{-0.06}$& -- & --\\
14098& 2.68$^{+0.12}_{-0.13}$& --&--\\
14099& 0.73$^{+0.36}_{-0.38}$ & --& --\\
14100& 0.43$^{+0.05}_{-0.06}$ & -- & --\\
14101& 0.30$^{+0.06}_{-0.07}$& ---& ---\\
14102& 2.02$^{+0.18}_{-0.20}$ & --- & ---\\
\hline
\end{tabular}

\end{table*}

\begin{figure*}
\includegraphics[scale=0.8]{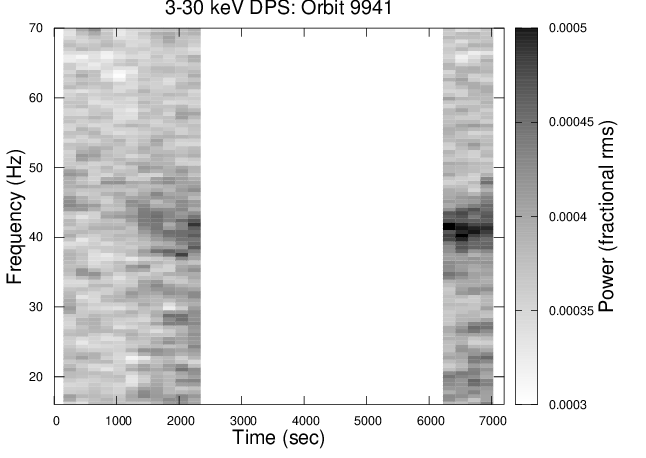}
\includegraphics[scale=0.8]{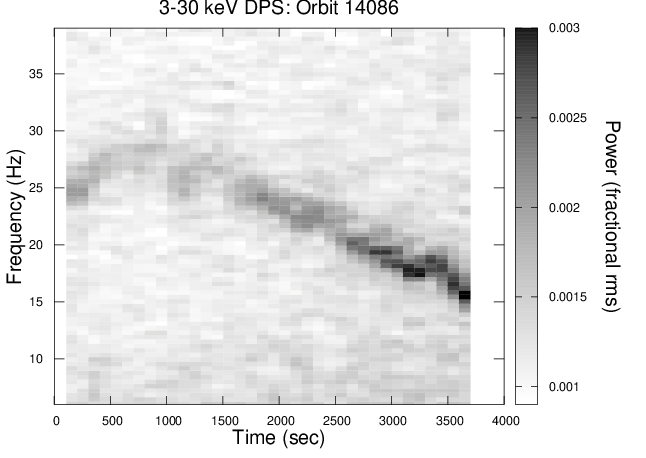}
\includegraphics[scale=0.8]{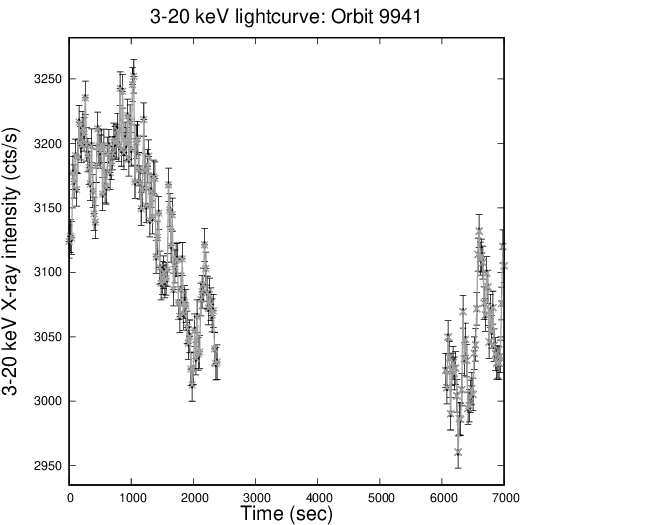}
\includegraphics[scale=0.8]{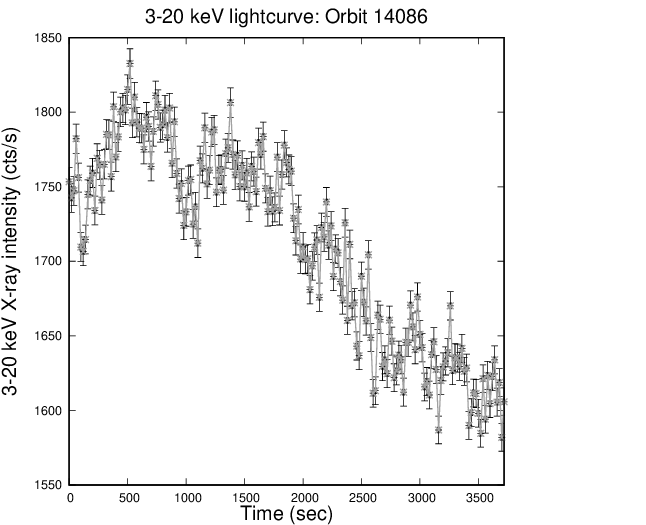}
\caption{Dynamic power spectra from {\it AstroSat}/LAXPC in the energy range 3-30~keV are shown in for two different observations: during HB/NB vertex (top left), and during HB (top right), and their corresponding lightcurves are shown the bottom left and bottom right panels respectively. The evolution of QPO during HB nearly follows the X-ray intensity pattern. However, no apparent correlation can be observed during HB/NB Vertex QPO at $\sim$41~Hz. }
\label{dps1}
\end{figure*}

\begin{figure*}
\includegraphics[scale=0.8]{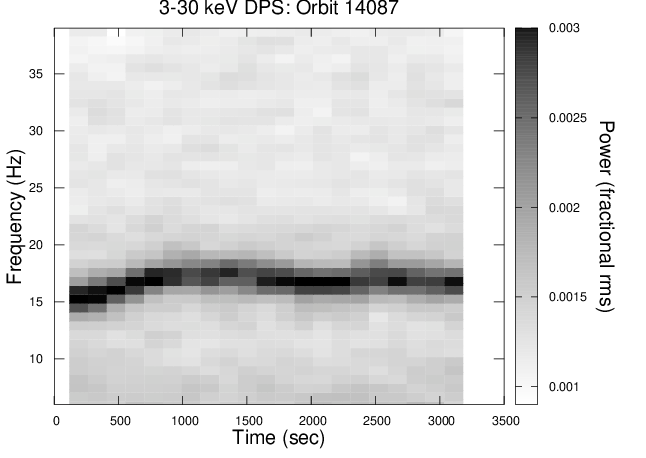}
\includegraphics[scale=0.8]{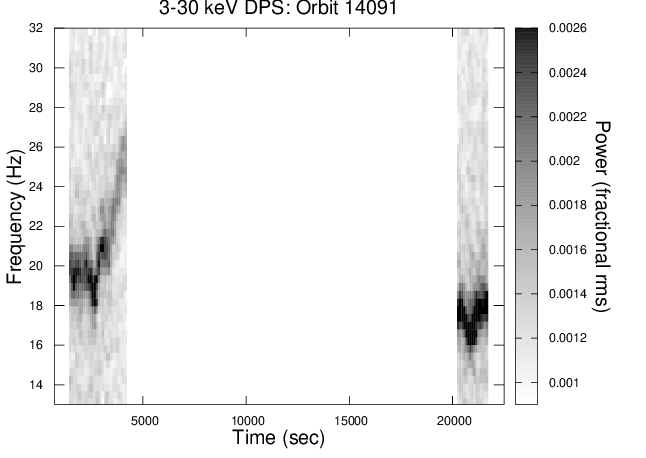}
\includegraphics[scale=0.8]{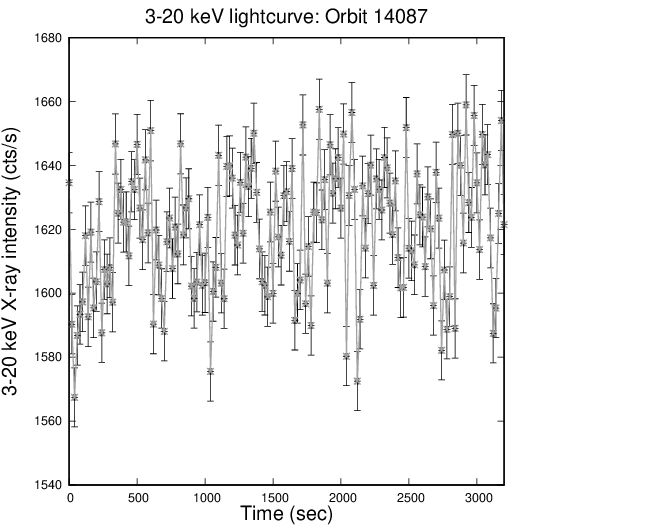}
\includegraphics[scale=0.8]{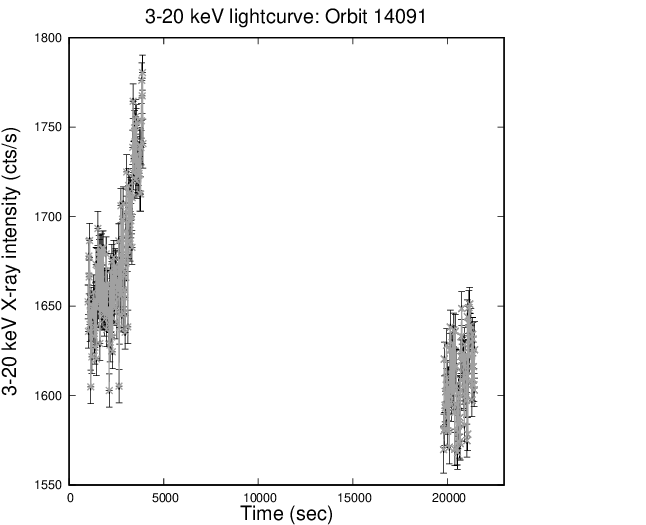}
\caption{Dynamic power spectra from {\it AstroSat}/LAXPC in the energy range 3-30~keV are shown in the top left and right panels for two different observations during HB, and their corresponding lightcurves are shown in the bottom left and bottom right panels, respectively. The QPO frequency during both cases follows the X-ray intensity pattern: either roughly constant over time or changes drastically.}
\label{dps2}
\end{figure*}

To understand whether the double-peaked QPOs are due to the rapidly changing QPO frequency, we have extracted dynamic power spectra (DPS) in the 3-30~keV energy range for orbits 9941, 14086, 14087, and 14091. For better visibility of DPS features, we used 1~Hz frequency bin size and 100~s time bin size to extract and plot DPS. The results are shown in the top panels of Figures~\ref{dps1} and \ref{dps2}. 3--20~keV lightcurves corresponding to the same time as DPS are shown in the bottom panels of both figures. The left panel of Figure~\ref{dps1} shows that the $\sim$42~Hz QPO frequency is stable over time while the X-ray intensity falls off significantly, indicating a possible uncorrelated behaviour between the QPO frequency and the X-ray intensity. On the other hand, the right panel of Figure~\ref{dps1} hints at a  correlation between changing QPO frequency and the X-ray intensity. Both panels show the strongest QPO rms power during the lowest X-ray intensity regimes.  
To verify the above, we have studied DPS and correlated X-ray intensity in two more observations shown in both panels of Figure~\ref{dps2}. The QPO frequency and the X-ray intensity observed during HBO at $\sim$16.9~Hz do not change with time, at least over a time scale of 3~ks and is shown in the left panel of Figure~\ref{dps2}. Double-peaked QPOs during orbit 14091 show a significant change in the QPO frequencies from $\sim$18~Hz to $\sim$26~Hz for a time scale of $\sim$3.5 ks and shown in the top right panel of Figure~\ref{dps2}. Again, a correlated change in X-ray intensity is also observed for the same time scale, and QPOs are strongest during the lowest X-ray intensity. Therefore, a possible correlation between X-ray intensity and QPO frequency exists in HBOs in the range of 15--30~Hz, while such correlation is not obvious for $\sim$42~Hz QPO, which occurs at the HB/NB vertex.

\begin{figure*}
\centering\includegraphics[scale=0.6,angle=-90]{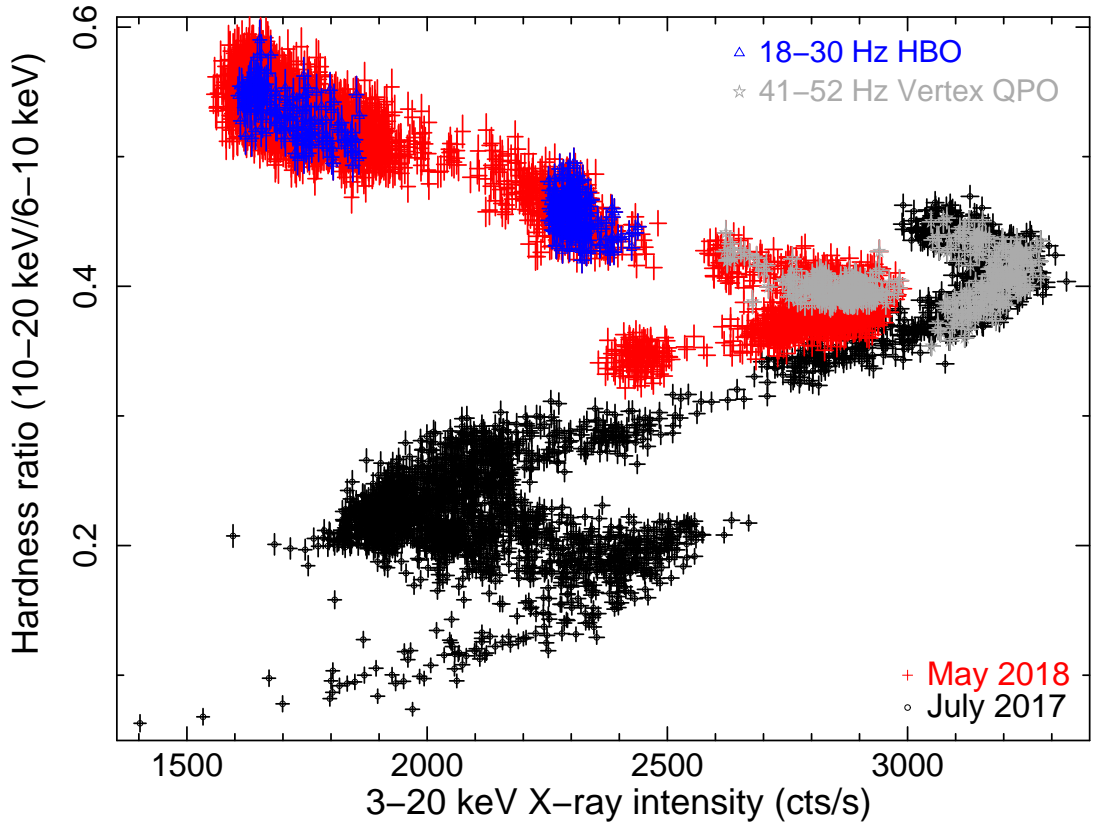}
\caption{Occurrences of different $>$ 10~Hz features in the PDS are shown in the Hardness intensity diagram using LAXPC. The HID from the observation in 2017 is shown with circles, and in 2018 is shown with pluses. The definition of hardness ratio and intensity are the same as Figure~\ref{light-hid-2017} and Figure~\ref{light-hid-2018}. HBOs in the range of 15--30~Hz are shown in triangles, while higher frequency QPOs in the range of 41--52~Hz are shown by stars. During both epochs, HB/NB vertices show the highest frequency HBOs observed in GX~340+0.}
\label{hid-qpo}
\end{figure*}

For a complete trace of the evolution of the  $>$10~Hz feature along the `Z' track, we combine the HID of 2017 and epoch together in Figure~\ref{hid-qpo} where the HID due to 2018 epoch is shown using crosses while circles show the HID of July 2017 epoch. A shift in HB/NB vertex between two epochs by $\sim$300 counts/s can be observed. Such a shift of the `Z' track in the HID was also seen earlier \citep{fr15}. The position of 16--30~Hz QPOs is shown by triangles over HID, while the position of 41--52~Hz QPOs is shown by stars. During both epochs, the highest frequency QPOs occupy the HB/NB vertex while 16--30~Hz QPOs fall on the horizontal branch. Therefore, as the source moves from HB towards HB/NB vertex, QPO frequency increases.   

\begin{figure*}
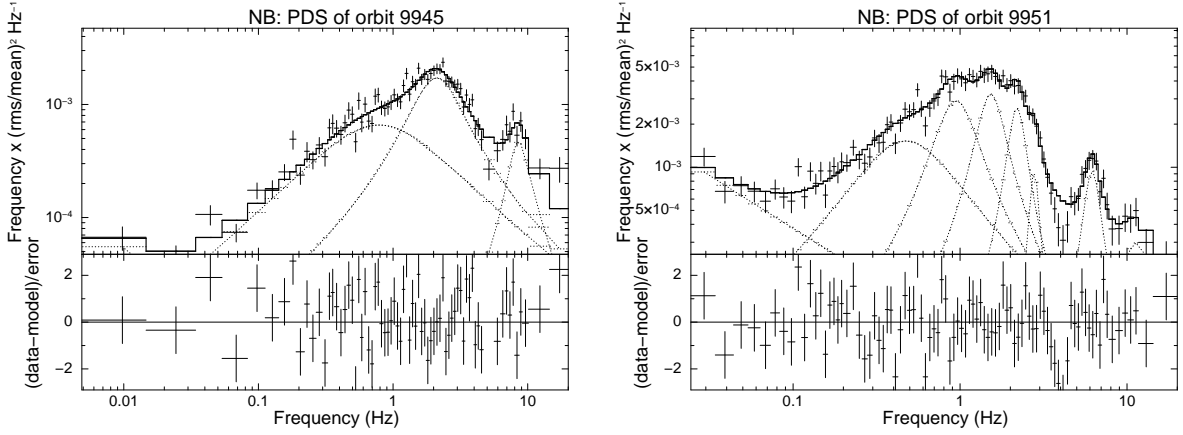

\includegraphics[scale=0.30,angle=-90]{fig7a.ps}
\includegraphics[scale=0.30,angle=-90]{fig7b.ps}
\caption{{\it \asat{}/LAXPC PDS} rms-normalised and Poisson noise-subtracted power density spectra (PDS) during Normal Branch are shown for orbit 9945 (left) and orbit 9951 (right) in 3--30~keV respectively. PDS are fitted with Lorentzians and power law, and the residuals are shown. During both orbits, PDS are dominated by a strong, hump-like noise component nearly $\sim$1--2Hz along with low-frequency QPO in 7--9~Hz. NBOs are detected above the 3$\sigma$ level.}
\label{nbo}
\end{figure*}

During the 2017 epoch that covers the entire normal branch, we have detected low-frequency QPOs ($<$10~Hz) known as normal branch oscillations (NBOs). Table~\ref{2017-pds-table} provides the details of NBOs detected during each orbit from the 2017 epoch. Two NBOs observed during orbit numbers 9945 and 9951 are shown in the left and right panels of Figure~\ref{nbo}. Two QPOs are detected over a significance of 5$\sigma$ at the frequency of $\sim$7.48~Hz and $\sim$6.21~Hz, respectively. A broad and strong Lorentzian noise component is observed around $\sim$2~Hz in all NB PDS. A comparison of Table~\ref{2017-pds-table} and Table~\ref{2018-pds-table} shows that NBOs are significantly weaker than HBOs.
We have not detected any flaring branch oscillations (FBO) during the 2017 epochs, where flaring branches are observed during orbit numbers 9947, 9953, 9955, and 9957, respectively. From Table~\ref{2017-pds-table}, we may note that the total rms amplitude of 0.05--70~Hz FB PDS is significantly lower (1.1--2.3\%) than that observed during NB (4--15\%). No QPO was detected in EFB, either. A 6~Hz QPO in the FB of GX~340+0 has been reported by \citet{se13} with an FWHM of 11.7~Hz.  

\section{Soft X-ray timing analysis}

\subsection{SXT}
\begin{figure*}
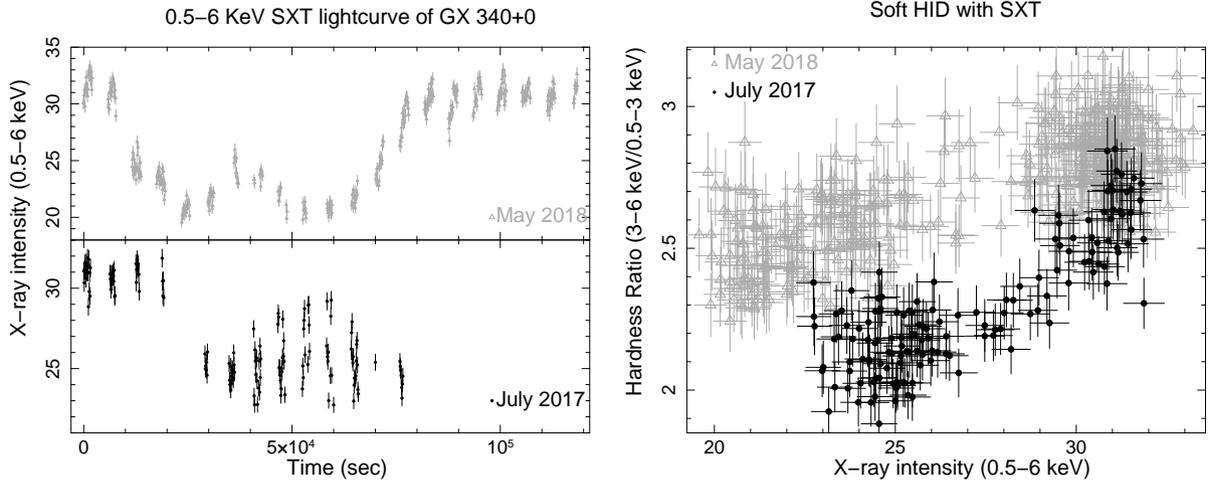

\centering
\includegraphics[scale=0.32,angle=-90]{fig8a.ps}
\includegraphics[scale=0.32,angle=-90]{fig8b.ps}
\caption{Soft X-ray behaviour of GX~340+0: \asat{}/SXT lightcurves extracted in the energy range 0.5--6.0~keV are shown for the May 2018 epoch (top left panel) and July 2017 (bottom left panel) epochs, respectively. A bin size of 100 s is used. The right panel shows a soft Hardness Intensity Diagram (HID) as a function of 0.5--6~keV X-ray count rate plotted using SXT observations. The soft hardness is defined as the ratio of the count rate in the energy range 3--6~keV and 0.5--3~keV, respectively. Horizontal, normal branches and HB/NB vertex are clearly visible.}
\label{sxt-hardness}
\end{figure*}

The soft band ($<$ 3~keV) behaviour of `Z' type neutron star X-ray binaries has not been explored due to the heavy pile-up issue in the detector at the focal plane in different observatories. However, SXT is an exception because of the large PSF size. Therefore, we first time attempted to study the `Z' track of GX~340+0 using SXT observations of the 2017 and 2018 epochs. Observation details are provided in Table~\ref{obs}. The left panel of Figure~\ref{sxt-hardness} shows 0.5--6~keV SXT lightcurves of the May 2018 (top left) and July 2017 (bottom left) epochs, respectively. The bin size used is 100~s to improve signal-to-noise without losing the long-term variability of the source. The long-term variabilities in both lightcurves are qualitatively similar to that observed in 3--20~keV LAXPC lightcurves shown in Figure~\ref{light-hid-2017} and Figure~\ref{light-hid-2018}. Comparing hard band LAXPC lightcurves (3--20~keV; left panels of Figure~\ref{light-hid-2017} and Figure~\ref{light-hid-2018}) with that of soft band SXT lightcurves, we notice that (1) A soft X-ray flux transition by a factor of $\sim$1.5 is observed in both lightcurves. The change in the flux follows the same pattern as observed from both epochs of LAXPC. (2) The flux transition during both epochs occurs simultaneously, as observed from LAXPC light curves. 

To identify whether such transitions during both epochs lead to the branch transition in the HID, we perform a soft HID analysis.
We have defined soft hardness as the ratio of count rate in 3--6~keV and 0.5--3~keV and plotted it as a function 0.3--6~keV. Such a definition ensures that soft X-ray emission exclusively dominates the HID. The right panel of Figure~\ref{sxt-hardness} shows the soft HID using data from both epochs. The horizontal branch, HB/NB vertex, normal branch, and NB/FB vertex are clearly visible in the HID. The soft band HB is distinctly different from the hard band HB observed in the right panel of Figure~\ref{light-hid-2018}: the soft HB prefers a positive slope in the HID, while the hard HB (as seen from the `Z' track of 2018 epoch) prefers a negative slope in the HID. For the first time, we present here the soft band `Z' track of GX~340+0. However, the flaring and extended flaring branches are not visible. Due to the consideration of the large bin size (100~s) for the HID and relatively lesser observational coverage of SXT compared to LAXPC, we have not clearly obtained the FB and EFB in the HID. However, the NB/FB vertex is observed. Previous studies showed that the EFB branch observed in `Z' sources during dips spanning 40--80~s \citep{2012A&A...546A..35C}. 

\subsection{\nicer{} }
\begin{figure*}
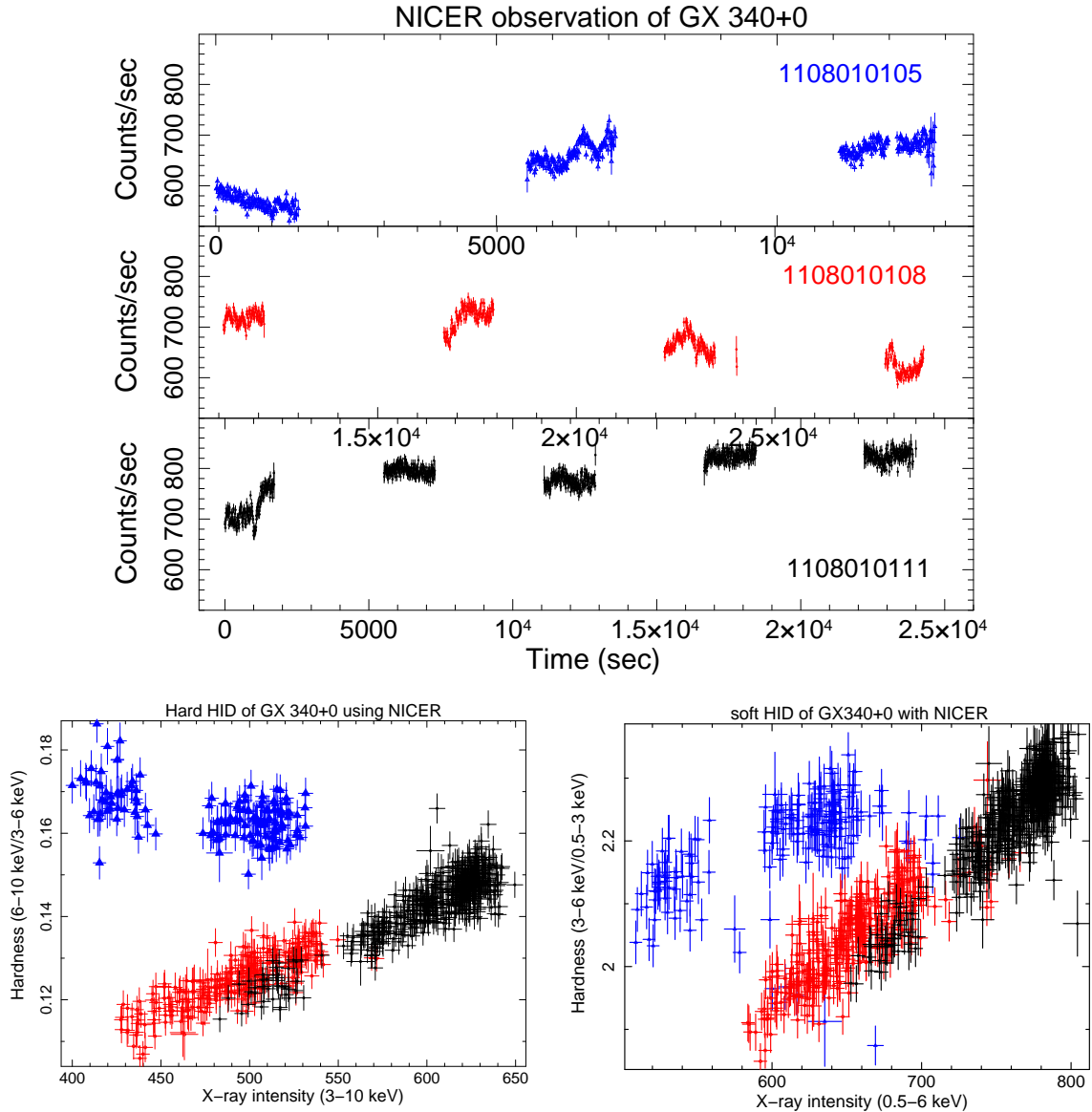

\centering
\includegraphics[scale=0.5,angle=-90]{fig9a.ps} 
\includegraphics[scale=0.3,angle=-90]{fig9b.ps}
\includegraphics[scale=0.3,angle=-90]{fig9c.ps}
\caption{{\it {\it NICER} view of GX~340+0} Top panel shows {\it NICER} lightcurves during three long orbits in 0.5--10~keV. The hardness intensity diagram (HID, constructed from the same light curves) is shown in the bottom panels. Hardness is defined as the count rate ratio in the energy range of 6--10~keV and 3--6~keV (bottom left panel). The horizontal branch (shown by triangles) and normal branch (shown by circles) are clearly visible in the HID. To compare the \nicer{} HID with that from SXT, we also extracted hardness using the same definition as used for SXT data and shown in the bottom right panel. A change in the HB slope is visible.}
\label{nicerhid}
\end{figure*}

\nicer{} observation details are provided in Table~\ref{obs}. Lightcurves of the three longest observations of GX~340+0 during the 2018 campaign are extracted in the energy range 0.5--10~keV and shown in three subpanels of the top panel in Figure~\ref{nicerhid}. Ranges of the y-axis are kept the same for comparison. A significant count rate variation is observed between 560 cts/s and 880 cts/s during three epochs. The bottom left panel of Figure~\ref{nicerhid} shows the hardness intensity diagram where the hardness is defined as the ratio of count rate between 6--10~keV and 3--6~keV, and the X-ray intensity is defined as the count rate in 3--10~keV. The definition of HID is chosen so HB and NB can be clearly visible. Using the definition of hardness similar to that of SXT, we extract the HID using \nicer{} observations and show them in the right panel of Figure~\ref{nicerhid}. HID from SXT and \nicer{} shows a qualitatively similar shape where HB has a positive slope. Comparing \nicer{} HIDs from both panels of Figure~\ref{nicerhid} where soft and hard bands are used, we note that the HB shows a negative and positive slope during hard and soft HIDs, respectively. A negative slope has also been observed in the hard HID with LAXPC.
Other observation epochs during the 2018 campaign either have very short exposures or their count rate and hardness overlap with either of the three observations presented here. Therefore, we have not plotted or used them for any further analysis. A part of HB and NB can be seen from the \nicer{} HID.

\begin{figure*}
\centering
\includegraphics[width=17cm,height=15cm]{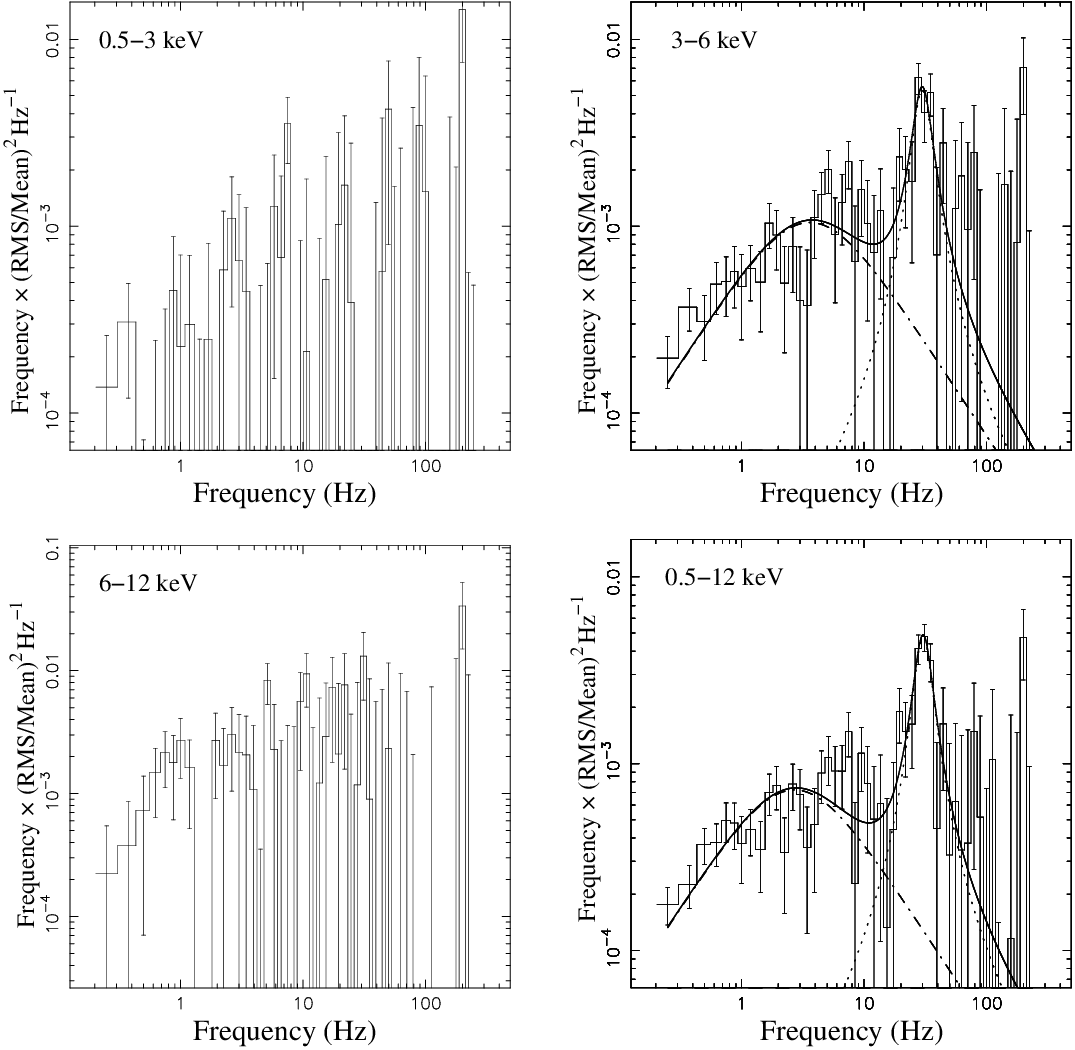}
\caption{{\it NICER energy-dependent power density spectra :} rms normalized, and Poisson noise subtracted power density spectra (PDS) are shown during the horizontal branch observations (triangles in Figure~\ref{nicerhid}) using {\it NICER}. QPOs due to horizontal branch oscillations (HBO) are detected at $\sim$33~Hz 3--6~keV (top right panel), while no QPOs are detected at 0.5--3~keV (top left panel) and 6--12~keV (bottom left panel). The broadband PDS in 0.5--12~keV is shown in the bottom right panel.}
\label{nicerpds}
\end{figure*}

For all three observation IDs, we have extracted rms-normalised and Poisson-noise subtracted PDS and fit each PDS with combinations of Lorentzian functions. A strong QPO is observed at 33.1 $\pm$ 1.1~Hz in 0.5--12~keV during the HB with a significance of 4.8$\sigma$. No features are observed from the other two observations during the NB. 
Best-fit PDS for the \nicer{} observation ID 1108010105 that shows the QPO feature is shown in Figure~\ref{nicerpds} using four different energy bands: 0.5--3~keV (top left panel), 3--6~keV (top right panel), 6--12~keV (bottom left panel) and 0.5--12~keV (bottom right panel) respectively. A partial overlap of two energy bands increases the signal-to-noise ratio of detected features in both bands. While no QPO is detected in 0.5--3~keV with a 1$\sigma$ rms upper limit of 2.9\%, strong QPO corresponding to the HBO is detected in 3--6~keV, and 0.5--12~keV with a significance $>3\sigma$. No QPOs are observed in 6--12~keV, possibly due to the low count rate. We are reporting the first such detection of HBO with \nicer{} from a `Z' type NSXB, indicating the importance of having a large area of \nicer{} and long exposures to detect faint signals.

\section{Energy-dependent variability of QPOs}
\begin{figure*}
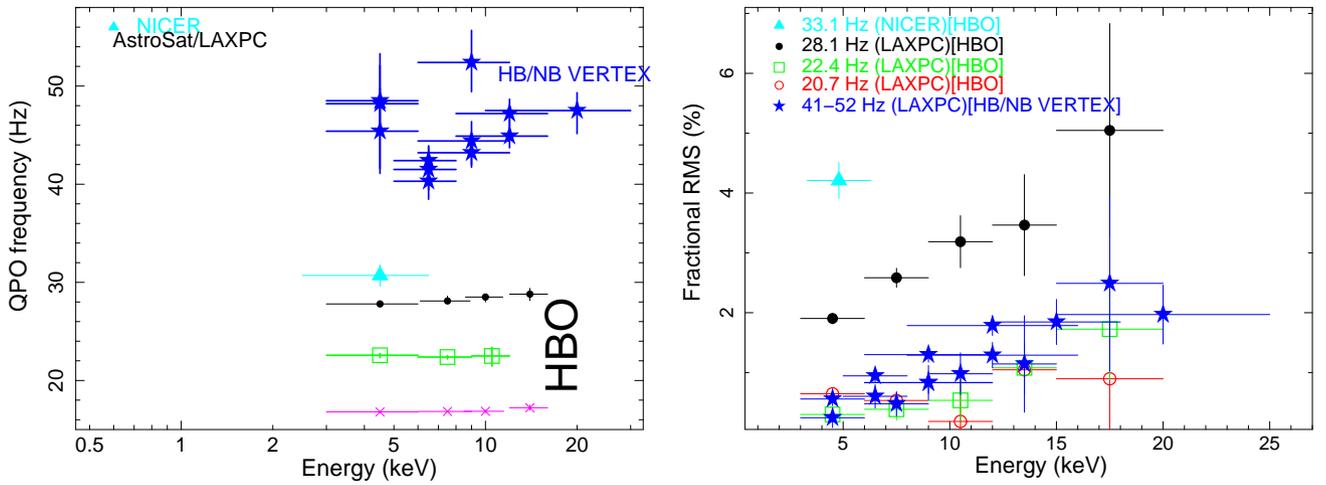

\includegraphics[scale=0.35,angle=-90]{fig11a.ps}
\includegraphics[scale=0.35,angle=-90]{fig11b.ps}
\caption{Frequencies of QPO due to the horizontal branch oscillations (HBO) and HB/NB vertex are shown as a function of photon energies in the left panel while their fractional rms as a function of photon energies are shown in the right panel. In both panels, \nicer{} observations are shown by triangles, while a similar colour convention is used for LAXPC observations except 16.9~Hz LAXPC QPO which is shown by magenta in the left panel. }
\label{qpo-rms-energy}
\end{figure*}

To understand the dependence of QPO features on photon energies and compare PDS properties of the same branch from two different missions, \asat{} and \nicer{}, we plot the QPO frequency and fractional rms as a function of photon energies in the left and right panels of Figure~\ref{qpo-rms-energy} respectively. A significant difference is observed between low-frequency HBOs (16--33~Hz) and high-frequency QPOs from HB/NB vertex (41--52~Hz). 
The HBOs observed in 16--33~Hz do not show a dependence of the QPO as a function of Energy and are consistent for both \nicer{} and \asat{} PDS analysis. However, 41--52~Hz QPOs during HB/NB apex, as observed from both epochs of \asat{}/LAXPC observations, show that the QPO frequency increases with the photon energies.

\begin{figure*}
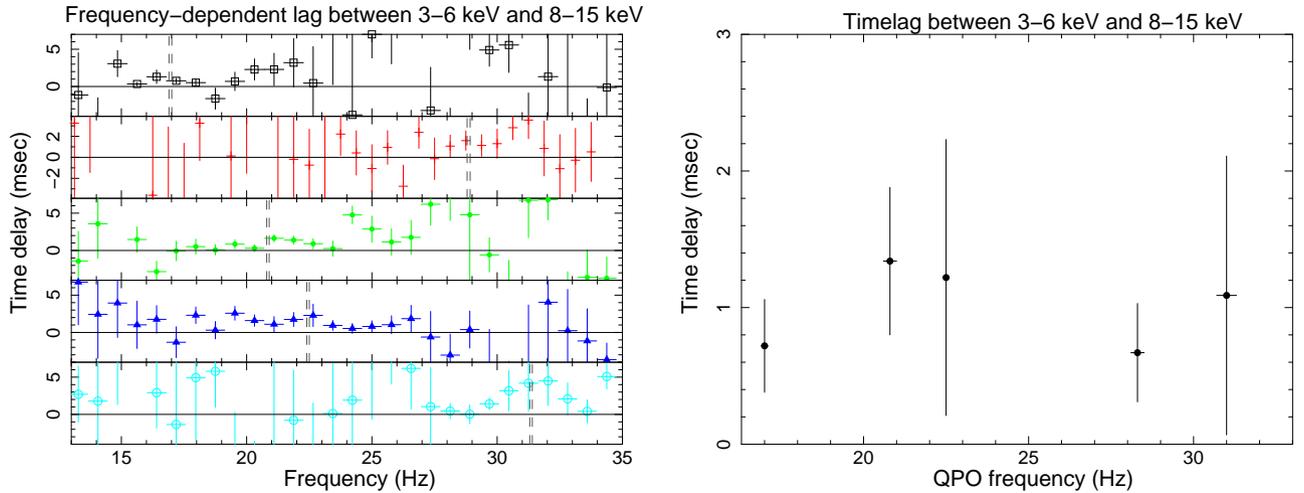

\includegraphics[scale=0.34,angle=-90]{fig12a.ps}
\includegraphics[scale=0.34,angle=-90]{fig12b.ps}
\caption{Left panel shows The Fourier-frequency dependent time lag between 3--6~keV and 8--15~keV for different orbits. Lag spectra obtained from orbit no. 14087, 14085, 14092, 14093, and 14095 are shown by open squares, crosses, closed circles, triangles, and open circles, respectively. In each subpanel, vertical double-dotted lines mark the position of QPOs. The right panel shows time lag between 3--6~keV and 8--15~keV as a function of QPO frequencies during HBO for the $\sim$17~Hz, $\sim$20.7~Hz, $\sim$22.4~Hz, $\sim$28.1~Hz, and $\sim$31.2~Hz QPOs, respectively. The observed lag where hard photons lagging soft photons shows a constant trend with the QPO frequencies. }
\label{qpo-lag}
\end{figure*}

At different QPO frequencies, fractional rms of the feature increase with the photon energy. This fact is consistent with previous studies \citep{jo00, bh23}. Although the rms-energy trend monotonously increases for HBOs from 16.9~Hz to 33.1~Hz from an average of 0.7\% $\pm$ 0.3 \% to 4.1\% $\pm$ 0.2. However, a sudden drop in the rms-energy trend is observed to a level of 1.3\% $\pm$ 0.3 during 41--52~Hz HB/NB apex QPOs.
While moving from HB to NB, such a dramatic transition in QPO properties is remarkable, possibly indicating they are of different origins \citep{sr11}. 
To understand the nature of HBOs observed at different frequencies, we computed a time lag between 3--6~keV and 8--15~keV at the different Fourier frequencies and shown in the left panel of Figure~\ref{qpo-lag} for five different orbits that showed QPOs between 17~Hz and 31~Hz. Except close to QPO frequencies, time-lag measurements are not significant in most parts of the frequency spectrum. For the same set of five observations, we calculate lag at different QPO frequencies of $\sim$17~Hz, $\sim$20.7Hz, $\sim$22.5~Hz, $\sim$28.1~Hz, and $\sim$31~Hz and shown in the right panel of Figure~\ref{qpo-lag}. The time lag between the hard and soft bands shows hard photons are delayed to soft photons by a few hundreds of microseconds to millisecond \citep[also seen in][]{bh23}; however, no specific trend is observed with QPO frequencies. The time lag for the same bands could not be constrained for the NBOs due to poor statistics and HB/NB Vertex QPOs as the QPO frequency changes with energy. 

\section{Discussion and conclusions}

\subsection{HID in hard and soft X-rays}
Observations of GX~340+0 with \asat{} in two different epochs covering effective exposures of $\sim$93~ks and $\sim$112~ks reveal the complete shape of the `Z' track in the HID from the HB to the dipping flaring branch (figures~\ref{light-hid-2017} and \ref{light-hid-2018}). The definition of hardness is so chosen that it is consistent with that from \citet{bh23}. A similar track has been observed previously by \xte{} \citep{jo00} with a slightly different definition of hardness. Interestingly, when we replaced the hard colour in the HID (the ratio of the count rate in 10.0--20.0~keV and 6.0--10.0~keV) with the soft colour (the ratio of count rate in 3.0--6.0~keV and 0.5--3~keV) using simultaneous \asat{}/SXT observations, we observed a similar `Z' shaped track (Figure~\ref{sxt-hardness}). The choice of energy ranges for soft band hardness is exclusive to that of conventional hardness. The \nicer{} observations in a similar period also show an HID similar to the one seen by SXT (Figure~\ref{nicerhid}).  The only difference between the soft and hard `Z' tracks we notice is the slope of the HB. Conventionally, it is flat or has a negative slope in Cyg~X-2 like Z-sources \citep{1989A&A...225...79H}, which implies softer when brighter behaviour. With SXT and NICER using a hardness ratio of 3--6~keV to 0.5--3~keV, we observe a positive slope of the HB. 
Such behaviour indicates that the component active in the 6--20~keV (causing a negative slope in the HB) has a lower contribution in softer X-rays (i.e. 0.5--6~keV). Additionally, there might be the presence of a soft component in SXT energy bands \citep[e.g. truncated accretion disk;][]{bh23}, which could also cause a change in the slope of the HB. GX~340+0 is also reported to have a strong intrinsic absorption \citep{ia06,ch06, se13,bh23}, which is also expected to change the shape of the track in the softer bands. Subsequent work will present a detailed `Z' track-resolved broadband spectral analysis of GX~340+0. 

Apart from HB, a similar trend in HID for both soft and hard bands implies that the underlying physical processes that drive the spectral evolution of the `Z' track in the hard band ($>$ 6~keV) also drive the spectral evolution in the soft band (0.5--6~keV). Thus, a similar set of components may be able to explain NB, FB and EFB evolution, albeit with the inclusion of a strong absorption column as necessary. Previously, using \nicer{} observations, atoll tracks have been observed in the soft HID (the soft colour is defined as the ratio of 1.1--2.0~keV and 0.5--1.1~keV) from an Atoll type NSXB 4U 0614+09 \citep{bu18}. 

\subsection{Temporal features in GX~340+0}
In our analysis of the PDS of the source during various \asat{} orbits, we observe QPOs at various positions on the HB and NB of the Z-track (see Figures \ref{hfpds},  \ref{dps1}, \ref{dps2}, \& \ref{nbo} and Tables~\ref{2017-pds-table} \& \ref{2018-pds-table}). These QPOs are observed in 16--33~Hz in the HB, 40--50~Hz near the upper vertex and 6.13--8.25~Hz in NB. We don't observe any statistically significant QPO in FB or EFB, possibly due to low exposure. Using \nicer{} observations (Figure~\ref{nicerpds}) for the first time, we report a statistically significant HBO in soft X-rays. The QPO is statistically detected in 0.5--4.5~keV, 3--6~keV and 0.5--12~keV energy ranges. The QPO in 6--12~keV is not detected, possibly due to low signal-to-noise (due to lower count rate in \nicer{} in that band). 
The HBO frequency is closely linked to the position on the HID, and in the \asat{} orbits where the source is observed to travel along the branch, the HBO is also observed to evolve accordingly. For example, in orbit 14086, the source is observed to have a significant reduction in count rate (see Figure~\ref{light-hid-2018}), and the PDS of that particular orbit shows multiple peaks (Figure~\ref{hfpds} top right panel). On investigating the DPS along with correlated X-ray intensity, the evolution of the timing feature becomes clear (Figure~\ref{dps1} \& \ref{dps2}). A similar evolution of the HB position and the QPO is also seen in the literature \citep[e.g. ][]{jo00, bh23}. 
A comparison of HBO properties from \asat{} and \nicer{} shows that in the frequency range 16--33~Hz, HBOs show a similar trend, e.g., fractional rms of the feature increasing as a function of the QPO (see Figure~\ref{qpo-rms-energy} right panel). However, the rms of the QPO near the upper vertex is quite lower than the oscillations in the HB. The trend of the HBO rms as a function of energy indicates that the feature is arising from a component that is stronger in harder X-rays. \citet{bh23} suggests that the Comptonising medium is strongly associated with the QPO which is further enforced by the non-detection of the QPO in 0.5--3~keV (which is dominated by the thermal components) even though the average background-subtracted count rate is significantly high in 0.5--3~keV (211 $\pm$ 0.8) and 3--6~keV (431 $\pm$ 0.6) energy bands.
Non-detection of \nicer{} HBO in 6--12~keV is solely due to the poor counting statistics (average count rate of 56 $\pm$ 2).

The QPO frequencies in the 16--33~Hz range are mostly independent of photon energies, while 41--52~Hz HB/NB apex QPOs show a significant change in QPO frequency with photon energies (Figure~\ref{qpo-rms-energy} left panel). These two differences may imply that 41--52~Hz HB/NB apex QPOs may have different origins than conventional HBOs. Since such QPOs have never been observed in any other part of the `Z' track, they are transient and may be strictly associated with the transition from the HB to the NB. Spectral properties of HB/NB vertex, which are shown in Figure 3 from \citet{bh23}, imply a very flat power-law index, high optical depth, and highest flux. Combining spectral properties with the highest frequency of the observed QPOs from HB, we may conclude that HB/NB vertex QPOs may have originated very close to the neutron star surface and are possibly associated with the optically thick boundary layer.  

The time lag computed between the soft band (3--6~keV) and the hard band (8--15~keV) and at the QPO frequency shows a hard lag, i.e. hard photons are delayed with respect to the soft photons (Figure~\ref{qpo-lag}). The lag between these bands could represent a delay introduced in the process of Comptonisation as the optical depth associated with the non-thermal component is quite high in this system \citep{bh23}. Our lag measurements are 
consistent with previous studies: \citet{de13} measured the lags of the upper kHz QPO in 4U 1608$-$52 and 4U 1636$-$53 shows that they are hard (hard photons lagging soft photons) and that their magnitude is independent of frequency and energy. However, \citet{pe22} observed 33 $\pm$ 35 $\mu$s lag at the kHz QPO during the `Z' phase of XTE J1701-462.

It is difficult to interpret the lag at QPO frequencies in the range of $\sim$41--52~Hz since the frequency changes with energy, and thus, it becomes difficult to correlate the variations in different energy bands.
We observe the NBOs throughout the segments where the source was present on NB. Properties of NBOs are consistent with previous studies \citep{jo00}. The NBO is also accompanied by a low-frequency broad noise component, which is weaker than the one observed in the HB. The rms of the NBO are also observed to be weaker than the HBO, but the dependence of the rms on the energy is similar. Lag values computed between 3--6~keV and 8--15~keV are not significant at NBO frequencies.

\section{Summary}
Here, we summarise the main results from the timing analysis of GX~340+0 using \asat{} and \nicer{} observations.
\begin{enumerate}
    \item We investigate the complete Z-track using the \asat{}/LAXPC observations and study the temporal properties of the source as a function of time. 
    \item For the first time, we depict the HID of the GX~340+0 in soft X-rays and demonstrate the significant change in the shape of the track as we study the source in soft X-rays
    \item For the first time, \nicer{} has detected the HBO in the 3--6~keV energy range around 32~Hz.
    \item Comptonization lag of the order of a few hundred microseconds is detected in QPO frequencies ranging from 6 to 31~Hz.
    \end{enumerate}
\section*{Acknowledgement}  
The present work makes use of data from the \asat{} mission of the Indian Space Research Organisation (ISRO), archived at the Indian Space Science Data Centre (ISSDC). This work has been performed utilizing the calibration databases and auxiliary analysis tools developed, maintained, and distributed by \asat{}/LAXPC teams with members from various institutions in India and abroad. This work has used the data from the Soft X-ray Telescope (SXT) developed at TIFR, Mumbai, and the SXT POC at TIFR is thanked for verifying and releasing the data via the ISSDC data archive and providing the necessary software tools. This work was supported by NASA through the NICER mission and the Astrophysics Explorers Program. 
This research has made use of data and/or software provided by the High Energy Astrophysics Science Archive Research Center (HEASARC), which is a service of the Astrophysics Science Division at NASA/GSFC.
AW gratefully acknowledges support from the Royal Astronomical Society in the form of a Summer Undergraduate Bursary and from the University of Southampton in the form of an Excel Summer Studentship.
LZ acknowledges the support of the National Natural Science Foundation of China (NSFC) under grant 12203052.

\section*{Data Availability}
The AstroSat data can be downloaded from  \url{https://astrobrowse.issdc.gov.in/astro archive} using the observation IDs mentioned in Table~\ref{obs}. The \nicer{} can be downloaded from the HEASARC data archive using the observation IDs mentioned in Table~\ref{obs}.

\bsp

\label{lastpage}

\end{document}